\documentclass[aps,prl,reprint,superscriptaddress,showkeys,longbibliography]{revtex4-2} 

\usepackage[T1]{fontenc} 
\usepackage{lmodern}

\usepackage{amsmath,amssymb}
\usepackage{mleftright} 
\usepackage{graphicx}
\graphicspath{{../figures/}}

\usepackage{dcolumn}
\usepackage{dsfont}
\usepackage{mathtools} 
\usepackage{subfigure}

\usepackage{url,hyperref}
\usepackage[table,usenames,dvipsnames]{xcolor}
\hypersetup{colorlinks=true, linkcolor=BrickRed,
  urlcolor=blue!50!black, citecolor=blue!50!black}

\usepackage[capitalise]{cleveref}
\crefrangelabelformat{equation}{(#3#1#4)$-$(#5#2#6)}

\usepackage{xfrac}    
\usepackage[nice]{nicefrac} 

\usepackage{siunitx}

\renewcommand\epsilon{\varepsilon}
\renewcommand\phi{\varphi}
\renewcommand\theta{\vartheta}
\renewcommand\rho{\varrho}
\renewcommand\vec[1]{{\boldsymbol #1}}
\newcommand\unitvec[1]{\vec{#1}}

\newcommand\diff{\mathrm{d}}

\newcommand\e{\text{e}}
\renewcommand\i{\text{i}}
\renewcommand\geq\geqslant
\renewcommand\leq\leqslant

\DeclareMathOperator\Imag{Im}

\newcommand\fc{f_\text{c}}
\newcommand\fL{f_\text{L}}
\newcommand\fA{f_\text{A}}
\newcommand\vD{v_\text{D}}
\newcommand\vDa{\vD^{(\infty)}}
\newcommand\vDp{\vD^\text{(p)}}
\newcommand\vL{v_\text{L}}
\newcommand\vA{v_\text{A}}
\newcommand\DR{D_\text{R}}
\newcommand\Deff{D_\text{eff}}
\newcommand\Dfree{D_\text{free}}
\newcommand\tauL{\tau_\text{L}}
\newcommand\tauR{\tau_\text{R}}
\newcommand\uxc{u_{x,\text{c}}}

\begin{document}

\title{Depinning Transition of Self-Propelled Particles}

\author{Arthur V. Straube}%
\email{straube@zib.de}
\affiliation{
Zuse Institute Berlin, Takustra{\ss}e 7, 14195 Berlin, Germany}
\affiliation{Freie Universit{\"a}t Berlin, Department of Mathematics and Computer Science, Arnimallee 6, 14195 Berlin, Germany}

\author{Felix H{\"o}f{}ling}%
\affiliation{Freie Universit{\"a}t Berlin, Department of Mathematics and Computer Science, Arnimallee 6, 14195 Berlin, Germany}
\affiliation{
Zuse Institute Berlin, Takustra{\ss}e 7, 14195 Berlin, Germany}

\begin{abstract} 
Depinning transitions occur when a threshold force must be applied to drive an otherwise immobile system.
For the depinning of colloidal particles from a corrugated landscape,
we show how active noise due to self-propulsion impacts the nature of this transition,
depending on the speed and the dimensionality $d$ of rotational Brownian motion:
the drift velocity exhibits the critical exponent 1/2 for quickly reorienting particles, which changes to $d/2$ for slow ones;
in between these limits, the drift varies superexponentially.
Different giant diffusion phenomena emerge in the two regimes.
Our predictions extend to systems with a saddle-node bifurcation in the presence of a bounded noise.
Moreover, our findings suggest that nonlinear responses are a sensitive probe of nonequilibrium behavior in active matter.
\end{abstract}


\maketitle


A depinning transition occurs when a physical system is driven out of an immobile, localized state by an external force $f$ such that, upon increasing the force above a critical value $\fc$, the system depins and starts to slide with a drift velocity $\vD$ \cite{Fisher:PR1998, Brazovskii-Nattermann:AP2004, Reichhardt-etal:RPP2017}.
When approaching the transition from above, this response to the driving varies as a power law,
$\vD\sim(f-\fc)^\beta$, with a universal scaling exponent $\beta$.
The phenomenon appears in a variety of contexts: it governs the onset of motion of
fronts \cite{Haudin-etal:PRL2009, Carpio-Bonilla:PRL2001, Martinez-Pedrero-etal:SR2016},
contact lines \cite{Jiang-etal:PRL2020}, and
domain walls \cite{Franke-etal:PRX2015, Woo-etal:NP2017, Bauer-etal:PRM2022}, but also of
vortices in superconductors \cite{Blatter-etal:RMP1994, Buchacek-etal:PRB2019, Fily:PRB2010}
and magnetic skyrmions \cite{Lin-etal:PRB2013}.
The depinning transition is fundamental for the phenomena of sliding friction and superlubricity \cite{Vanossi-etal-RMP2013, Bylinskii-etal:S2015, Vanossi-etal:NC2020,
Brazda-etal:PRX2018, Hod-etal:N2018},
synchronization \cite{Juniper-etal:NC2015},
and locking \cite{Korda-etal:PRL2002, Balvin-etal:PRL2009, Juniper:NJP2017, Cao:NP2019, Stoop-etal:PRL2020, Chepizhko:PRL2022}.
Colloidal systems have given exquisite insight into the depinning transition of
individual particles \cite{Evstigneev-etal:PRE2008, Straube-Tierno:EPL2013, Ma-etal:SM2015, Juniper-etal:PRE2016},
monolayers \cite{Pertsinidis-Ling:PRL2008, Bohlein-etal:NM2011, Tierno:PRL2012,
Huelsberg-Klapp:PRE2023},
and in glasses \cite{Winter:JCP2013,Senbil:PRL2019,Gruber:PRE2020}.

Unlike passive matter, active particles---motile microorganisms, artificial microswimmers, and active colloids---propel themselves and perform a directed motion,
with the direction randomized as a function of time
\cite{Romanczuk:EPJST2012,Aranson:PU2013,Elgeti:RPP2015,Bechinger:RMP2016,Zoettl:JPCM2016}.
Experimental research in the field is fueled by the vision of microrobots performing
specific transport tasks \cite{Nelson:ARBE2010,Palagi:NRM2018,Alapan:SR2018};
such particles move through structured channels, blood vessels, or surmount geometric constrictions \cite{Xiao-etal:AMI2019}.
More fundamentally, the inherently nonequilibrium nature of self-propulsion leads to a nontrivial interplay
with a patterned substrate \cite{Zoettl:JPCM2016,RezaShaebani:NRP2020,Choudhury:NJP2017,Straube:PRR2019,Ryabov:SR2023},
with impact on the macroscopic transport and inducing, e.g., directionality \cite{Bag:JPCL2022},
negative mobility \cite{Ghosh:PRE2014, Rizkallah:PRL2023},
or superdiffusion \cite{Pattanayak:EPJE2019}.
One anticipates that self-propulsion also has significant ramifications on the depinning transition, which is an open issue.

In this work, we answer this basic question within the paradigm of the active Brownian particle (ABP) driven over a periodic landscape.
The response is contrasted to that of a passive particle, whose drift velocity is known to exibit a power law with exponent $\beta=1/2$ near depinning.
We show that the activity
modifies the nature of the transition, including a change of the exponent to some $\beta'$, superexponential behavior, the emergence of another singular point, and an unbounded enhancement of the diffusivity in between.
The new exponent $\beta'=d/2$ is sensitive to the dimensionality $d$ of rotational Brownian motion.

\paragraph{Model.}

\begin{figure*}
	\includegraphics[width=1.0\textwidth]{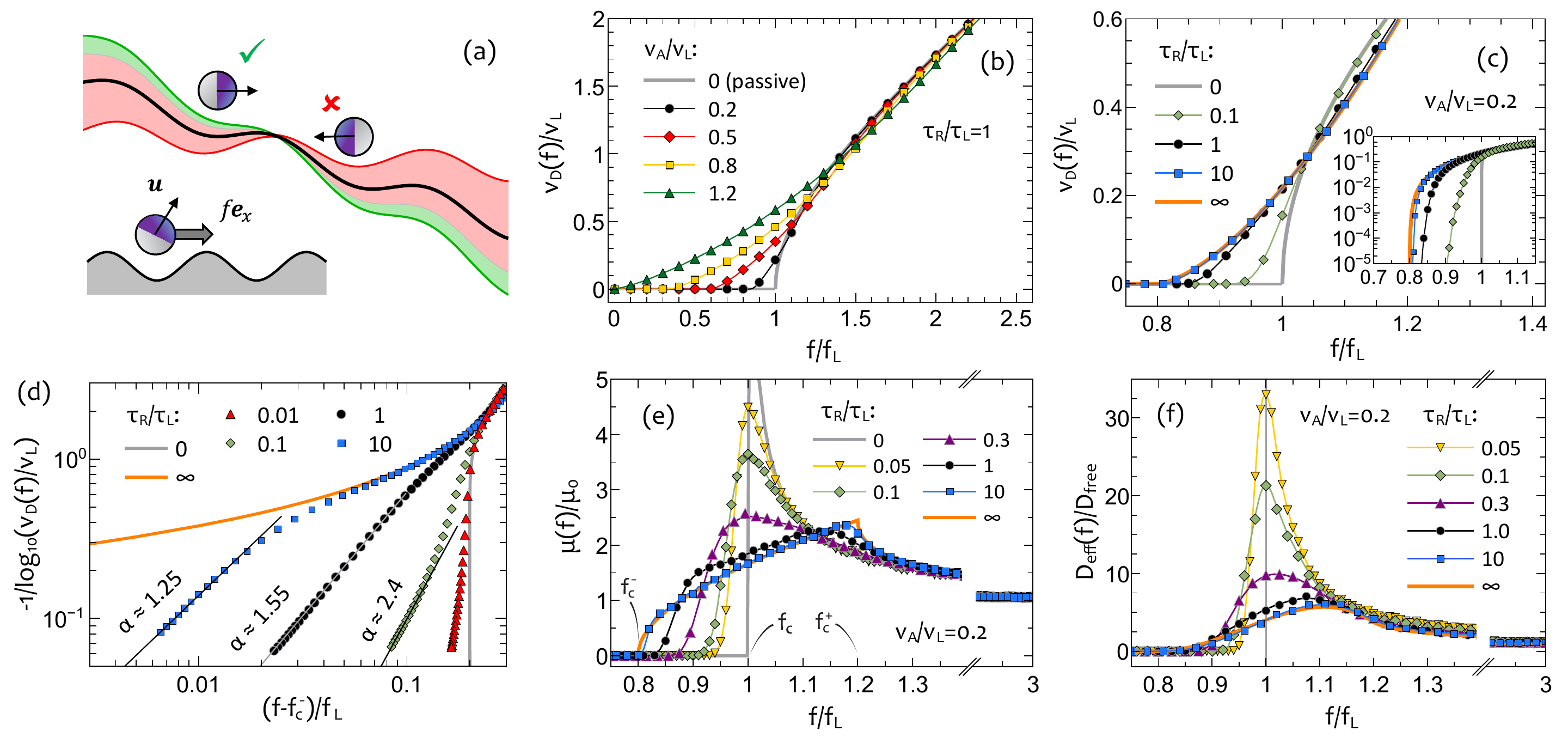}
	\caption{
	Panel~(a): Depinning of an active Janus particle from a corrugated potential landscape and subject to a driving force $f$ (bottom) and the mapping to passive motion in a randomly tilted potential landscape (top). The tilt has a constant contribution $-f x$ (black line), which is increased (green line) or decreased (red line) depending on the orientation $\vec u(t)$.
	Panels~(b),(c): Drift velocity $\vD(f)$ of the active particle with (b)~fixed rotational persistence time $\tauR=\tauL$ but varying propulsion velocity $\vA$ and (c)~fixed $\vA = 0.2 \vL$ but varying $\tauR$.
  In panel~(b), the inset shows the same data as in the main panel on a logarithmic scale.
  Panel~(d): high-precision data for $\vD(f)/\vL$ from the numerical solution of the Fokker--Planck equation, shown on an iterated logarithmic scale and corroborating the superexponential convergence of $\vD(f) \to 0$ as $f \downarrow \fc^-$ [\cref{eq:superexp}].
	Panels~(e),(f): Differential mobility $\mu(f)=\diff \vD(f) / \diff f$ and effective diffusion constant
    $\Deff(f)$ as functions of the driving force $f$ for fixed $\vA = 0.2 \vL$ and varying $\tauR$.
	All panels: thin lines interpolate between stochastic simulation results (symbols);
	thick lines are analytic predictions for the limits of the hyper wobbler (gray, $\tauR \to 0$)
	and the lazy wobbler (orange, $\tauR \to \infty$).
  }
	\label{fig:combo}
\end{figure*}

We use the framework of an ABP confined to a periodic potential energy landscape $U(\vec r)$ and subjected to an external force $\vec f$ [\cref{fig:combo}(a)]. The position $\vec r$ and the orientation $\unitvec u$ of the ABP satisfy the It{\=o}--Langevin equations \cite{Leitmann:PRL2016, *Leitmann:PRE2017, Hoefling:2024}:
\begin{align}
\dot{\vec r}(t) & = \mu_0 [\vec f -\nabla U\boldsymbol(\vec r(t)\boldsymbol)] + \vA \vec u(t) + \vec\xi(t)\,, \label{eqn:LE-gen-r}  \\
\dot{\vec u}(t) & =  \vec\omega(t) \times \vec u(t) - \tauR^{-1} \vec u(t)\,,
\label{eqn:LE-gen-u}
\end{align}
where $\vA \geq 0$ is the propulsion strength and $\mu_0$ is the mobility of the free particle.
The random linear and angular velocities, $\vec\xi$ and $\vec\omega$, respectively, are Gaussian white noise processes with zero means, $\langle \vec\xi(t)\rangle =0$ and $\langle \vec\omega(t)\rangle =0$, and covariances
$\langle \vec\xi(t) \otimes \vec\xi(t') \rangle =2 D_0 \mathds{1} \delta(t-t')$ and
$\langle \vec\omega(t) \otimes \vec\omega(t') \rangle =2 \DR\mathds{1}\delta(t-t')$.
Here, $D_0$ and $\DR$ are the translational and rotational diffusion constants, respectively.
With $\tauR^{-1}=(d-1)\DR$, \cref{eqn:LE-gen-u} implies that
$\vec u(t)$ performs an unbiased diffusion on the unit circle ($d=2$) or unit sphere ($d=3$) \cite{Hoefling:2024}.
In the stationary limit, all directions of $\vec u$ are equally probable and
the evolution of $\vec u(t)$ yields $\langle \vec u(t) \rangle = 0$ and
$\langle \vec u(t) \cdot \vec u(t') \rangle = \exp(-|t-t'|/\tauR)$;
hence, $\tauR$ is the persistence time of the orientation.
In the following, we will mainly consider the case $d=3$ because of its relevance for
applications \cite{Enculescu:PRL2011,Choudhury:NJP2017,Gibbs:L2019,Carrasco-Fadanelli:PRR2023}.

As the potential, we employ the prototypical one-dimensional corrugated landscape \cite{Wen:PRE2023} with a cosine shape: $U(\vec r)=U_\textrm{L}(1-\cos kx)$ with $x = \vec r \cdot \unitvec e_x$ and the unit vector $\unitvec e_x$ pointing perpendicular to the ripples;
$U_\textrm{L}$ is the amplitude of the landscape and $k$ its wavenumber, equivalently, $\lambda = 2\pi/k$ its wavelength.
Translational symmetry allows us to fix the direction of the force to $\vec f=f \unitvec e_x$ with $f \geq 0$.
Focusing on the depinning singularity, we switch off the translational Brownian noise ($D_0=0$), which is known to
mask the singular behavior at the critical point such that a rounded rather than sharp transition is observed \cite{Stratonovich:1967, Middleton:PRB1992, Kolton:PRE2020}.
With this, the model reduces to an Adler equation amended by an ``active noise'' $\vA u_x(t)= \vA \vec u(t)\cdot \unitvec e_x$:
\begin{equation}
\dot x(t) = \mu_0[f - \fL \sin (kx(t))] + \vA u_x(t)\,.
\label{eqn:LE-act1d}
\end{equation}
The characteristic force $\fL=U_\textrm{L} k$, the velocity $\vL=\mu_0 \fL$ and the timescale $\tauL = \lambda/\vL$ serve us as a system of independent units.
Regimes of different responses are distinguished by the relative strengths of the external driving $f/\fL$, the active propulsion $\vA/\vL$, and the rotational noise $\tauR/\tauL$.
For the stochastic simulations, we combined Euler integration of \cref{eqn:LE-act1d} with a geometric scheme for \cref{eqn:LE-gen-u} \cite{Hoefling:2024} and noise reduction \cite{supplement}.
The drift velocity was calculated from averaging over the driven stationary ensemble as
$\vD(f) = \lim_{t \to \infty} \langle x(t) \rangle_f / t$;
the variance yielded the dispersion coefficient or effective diffusion constant:
$\Deff(f) = \lim_{t \to \infty} \mathrm{Var}[x(t)]_f / 2t$.

The r.h.s.\ of \cref{eqn:LE-act1d} may also be viewed as originating from a tilted potential,
$U(x) - [f + (\vA/\mu_0) u_x(t)] x$. Its barriers can only be crossed if $u_x > \uxc = (\vL/\vA)(1-f/\fL)$ [\cref{fig:combo}(a), green shading] and they act as a randomly rocking ratchet, rectifying the \emph{a priori} unbiased self-propelled motion and thus facilitating transport.

\paragraph{Depinning transition.}

For passive motion, $\vA=0$, the particle's response to the driving is governed by the dynamic system $\dot x = g(x,f)$ with $g(x,f) = \mu_0(f - \fL \sin kx)$, which exhibits a saddle-node bifurcation \cite{Strogatz:1994}. Two equilibria $x_*^{\pm} \in [0, \lambda)$, obeying $g(x_*,f)=0$, exist for $f < \fc$ and disappear at the critical point $\fc=\fL$,
which is determined by the additional requirement $\partial_x g(x_*,\fc)=0$.
Thus, the particle is pinned by the landscape for $f < \fL$ and remains immobile, $\vD(f)=0$.
For $f > \fL$, the particle slides with $\vD(f) = \lambda/\tau_1(f)$,
where $\tau_1(f) = \int_0^\lambda g(x,f)^{-1} \,\diff x$
is the time it takes the particle to travel one wavelength.
With the present potential, one finds for the drift velocity of the passive particle:
\begin{equation}
  \vDp(f) = \mu_0\sqrt{f^2 - \fL^2} \,, \qquad f > \fL,
  \label{eq:pass-depin}
\end{equation}
which admits for the scaling form $\vDp(f) = \vL s(f/\fL)$ with the rescaled force $y=f/\fL$ and the scaling function $s(y)=\sqrt{y^2-1}$ for $|y|>1$ and $s(y)=0$ otherwise.
Expanding \cref{eq:pass-depin} close to the critical point, $\fc=\fL$, shows that $\vDp(f)$ exhibits a square-root singularity,
\begin{equation}
  \vDp(f \downarrow \fc) \sim (f-\fc)^{\beta}, \quad  \beta=1/2.
  \label{eq:critical-passive}
\end{equation}

For the self-propelled particle, $\vA>0$, the force--velocity relationship obtained from the simulations shows
progressively stronger deviations of $\vD(f)$ from the square-root law \eqref{eq:pass-depin} upon gradually increasing the propulsion strength $\vA$ while fixing the orientational persistence time $\tauR=\tauL$ [\cref{fig:combo}(b)].
Conversely, changing $\tauR$ at fixed $\vA = 0.2 \vL$ yields a similar picture [\cref{fig:combo}(c)];
the dependencies remain the same qualitatively when using other values of $\tauR$ or $\vA$.
Importantly, the ABP with $\vA>0$ and $\tauR>0$ displays a nonzero drift also for $f<\fc$.
At first sight, this seems to resemble the rounding of the depinning transition caused by translational Brownian noise \cite{Stratonovich:1967}.
However, we will show that the effect of active propulsion on the transition is entirely different and cannot be mimicked by translational diffusion, $D_0 > 0$.
In particular, a pinned state exists in the presence of self-propulsion for $f < \fc^-$ with the new, shifted threshold $\fc^- = \fL - \vA/\mu_0$
\footnote{For fast active motion, $\vA \geq \vL$, the particle depins for any $f > 0$; we exclude this trivial situation here.}.

The existence of this activity-controlled critical force $\fc^-$ is justified by the second observation:
upon varying $\tauR$ from 0 to $\infty$ at fixed ratio $\vA/\vL < 1$, the force--velocity curves interpolate monotonically
between the analytical solutions for the ``hyper wobbler'' ($\vA > 0$, $\tauR \to 0$) and the ``lazy wobbler'' ($\tauR \to \infty$, i.e., $\DR \to 0$) [\cref{fig:combo}(c)].
The hyper wobbler is an ABP with a rapidly changing orientation such that $\tauR$ is the smallest timescale of the problem, $\tauR \ll \tauL$ and $\tauR \ll \tau_f = \lambda/(\mu_0 f)$.
Such an ABP quickly samples all possible orientations before any translation occurs and the active noise $u_x(t)$ is averaged out from \cref{eqn:LE-act1d}.
Thus, self-propulsion is inefficient for the hyper wobbler, which also obeys \cref{eq:pass-depin}.

In the opposite regime of a lazy wobbler ($\tauR \gg \tauL, \tau_f$), rotational motion is slow.
The trajectories $x(t)$ can be thought of as a one-dimensional random walk composed of a sequence of long independent segments $i=1,2,\dots$ with fixed orientations $\unitvec u_i$ isotropically distributed and randomly changing at random times with rate $\tauR^{-1}$.
The active noise term in \cref{eqn:LE-act1d} is specified by the polar angle $\theta \in[0,\pi]$ such that $u_x = \cos\theta$;
being constant here, the noise term can be absorbed in the shifted driving force $\fA(\theta)=f+(\vA/\mu_0)\cos\theta$.
With this, the dynamic system has the same form as above, $\dot x = g(x,\fA(\theta))$, and repeating the analysis leading to \cref{eq:pass-depin}, one arrives at $\vD(f;\theta)=\vL s(\fA(\theta)/\fL)$.
The velocity--force relationship of the lazy wobbler with prescribed orientation $\theta$ has the same functional form as for the passive particle [\cref{eq:pass-depin}].
Merely the condition $|\fA(\theta)| > \fc$ implies a shift of the critical point from $\fc=\fL$ to $\fL - (\vA/\mu_0)\cos\theta$.
The latter expression depends on $\theta$ and varies between the maximum and minimum values $\fc^\pm = \fL \pm  \vA /\mu_0$.
In particular, it holds $\vD(f;\theta) = 0$ for $f \leq \fc^-$ irrespectively of $\theta$.

At long times, the random walk implies a uniform average over the orientation,
$\langle \cdot \rangle_{\unitvec u} := (4\pi)^{-1} \int \cdot\, \sin\theta \,\diff\theta \,\diff \phi$,
and we find for the drift velocity
$\lim_{t\to\infty} \langle x(t) / t \rangle = \langle \vD(f;\theta)  \rangle_{\unitvec u} =: \vDa(f)$
of the lazy wobbler \cite{supplement}:
\begin{align}
  \vDa(f) = \frac{\vL^2}{2 \vA}
    \begin{cases}
    0, & f \leq \fc^-, \\
    w_{+}(f/\fL), & \fc^- < f <  \fc^+, \\
    w_{+}(f/\fL) - w_{-}(f/\fL), & f \geq \fc^+,
    \end{cases}
  \label{eq:active-depin}
\end{align}
introducing new scaling functions $w_{\pm}(y)= w(y \pm \vA/\vL)$ with
$w(z)=\int_1^z s(y) \diff y = [z s(z) - \ln(z + s(z))]/2$.
The passive limit [\cref{eq:pass-depin}] is recovered as $\vA \to 0$; in this limit, the two singular points $\fc^{\pm}$ converge to $\fc = \fL$.
Due to $w'(z) = s(z)$, the critical exponent $\beta$ increases by 1, turning the square-root singularity [\cref{eq:critical-passive}] into
\begin{equation}
  \vDa(f \downarrow \fc^-) \sim (f-\fc^-)^{\beta'}, \quad  \beta'=3/2 \,.
\end{equation}
The argument applies similarly for rotational motion in a plane, noting that $u_x$ is distributed differently in this case.
Analysis of the leading asymptotic behavior upon $\epsilon := (f - \fc^-) / \fL \downarrow 0$ yields for $d=2,3$ dimensions \cite{supplement}:
\begin{equation}
  \vDa(\epsilon \downarrow 0) \simeq \frac{\sqrt{d-1}}{d} \, \vL^{1/2 + \beta'} \vA^{1/2 - \beta'} \, \epsilon^{\beta'} \,.
 \label{eq:vDa-critdD}
\end{equation}
The new exponent $\beta'=d/2$ renders the appearance of $\vDa(f)$ near $\fc^-$ smoother than for a passive particle [\cref{fig:combo}(b,c)]; yet we stress that \cref{eq:active-depin,eq:vDa-critdD} predict a sharp transition.

Pictorially, the behavior of $\vDa(f)$ near $f\approx \fc^-$ may be understood from the random tilts of the potential landscape [\cref{fig:combo}(a)]: in an ensemble of particles, only those with orientations pointing sufficiently close towards the direction of the force contribute to the transport: $u_x > \uxc = 1 - \epsilon \vL/\vA$ so that $\vD(f;\theta) > 0$.
Near the transition, $\uxc \to 1$ and the square-root behavior $\vD(f;\theta) \sim (u_x - \uxc)^{1/2}$ is weighted with the distribution of $u_x$ close to 1; the latter is flat for $d=3$, but divergent $\sim (1-u_x)^{-1/2}$ for $d=2$.
Both factors combine into $\sim (1-\uxc)^{d/2}$ after integration and hence $\beta'=d/2$.
Transport near criticality is thus faster for $d=2$ than for $d=3$ (Fig.~S2 in \cite{supplement}).

\paragraph{Finite rotational diffusion.}

For $0 < \tauR < \infty$, away from the limiting cases, the polar angle $\theta(t)$ samples different orientations in the course of time. Regarding the transport, this kind of motion is less efficient than with a fixed orientation in the direction of the driving force ($\theta=0$), whereas the opposite direction ($\theta=\pi$) is
the most inefficient situation.
One concludes that the drift velocity $\vD(f)$ is bounded, $\vD(f;\theta=\pi)\leq \vD(f) \leq \vD(f;\theta=0)$;
in particular, $\vD(f) = 0$ for $f < \fc^-$ for all values of $\tauR$.

The drift velocity is furthermore bounded by the solutions for the passive particle and the lazy wobbler [\cref{eq:pass-depin,eq:active-depin}]:
$\vDp(f) \leq \vD(f) \leq \vDa(f)$ for all $\tauR$ and $f < f_x$ (\cref{fig:combo}c)
with $f_x$ being the force where the two bounds intersect.
For $f > f_x$, the bounds reverse their roles so that 
for strong driving, counterintuitively, active propulsion slows down transport compared to passive particles.

The described behavior of $\vD(f)$ is corroborated by precise numerical solutions of the corresponding Fokker--Planck equation, which allowed us to follow $\vD(f)/\vL$ down to \num{e-15} \cite{supplement}.
These semi-analytical results suggest a superexponential convergence to the critical point $\fc^{-}$,
\begin{equation}
 \vD(f) \simeq \vL \exp\mleft(-b (f-\fc^-)^{-\alpha}\mright)\,, \quad f \downarrow \fc^-;
 \label{eq:superexp}
\end{equation}
the coefficients $\alpha>1$ and $b>0$ depend on $\tauR$ and we found that $\alpha$ increases as $\tauR$ is decreased [\cref{fig:combo}(d)].
The form of \cref{eq:superexp} is in line with predictions from related discrete-time models \cite{Olicon-Mendes:PHD, Chigarev:CH2023}
and it is rooted in a very slow initial increase of the probability that the particle slips along $\unitvec e_x$ by one wavelength upon increasing $f > \fc^-$. (For $f < \fc^-$, this probability is zero.)
For $\tauR \gg \tauL$ and upon increasing $f$ further, the asymptotic behavior of $\vD(f)$ crosses over to closely follow the lazy-wobbler solution, $\vDa(f)$.
We conclude that $\vD(f) > 0$ for $f > \fc^-$, i.e., the critical point is the same for all $\tauR > 0$.

\paragraph{Differential mobility.}

The differential mobility $\mu(f) = \diff \vD(f)/\diff f$
may serve as an alternative measure of the transport which is more sensitive to singular behavior.
For finite $\tauR$, we have calculated $\mu(f)$ from the numerical results for $\vD(f)$, and
$\mu(f)$ is readily obtained for $\tauR \to 0$ and $\tauR \to \infty$ from \cref{eq:pass-depin,eq:active-depin}, respectively [\cref{fig:combo}(e)].
In any situation, the potential landscape becomes irrelevant for
sufficiently strong driving, $\mu(f\to\infty) = \mu_0$.
For the hyper wobbler ($\tauR \to 0$), the mobility diverges at the
corresponding critical force, $\mu_\text{p}(f\downarrow \fc) \sim (f-\fc)^{-1/2}$, whereas for the lazy wobbler it vanishes as
$\mu_\infty(f\downarrow \fc^-) \sim (f-\fc^-)^{1/2}$.
In addition, $\mu_\infty(f)$ remains finite but exhibits a cusp at $f=\fc^+$, pinpointing the presence of a second singular point, at which $\mu_\infty(f)$ is maximal.
In between these limiting cases, the mobility exhibits a maximum that, upon increasing $\tauR$, interpolates in peak height and position between the divergence at $f=\fc$ ($\tauR \ll \tauL$) and the cusp at $f=\fc^+$ ($\tauR \gg \tauL$).
Concomitantly, the left flank of the peak moves from $f=\fc$ to $\fc^-$, broadening the peak.

\paragraph{Activity-induced giant diffusion.}

For passive depinning, the differential mobility was found to be a good proxy of the dispersion coefficient, $\Deff(f) \propto \mu(f)$, which restores a linear response relation \cite{Costantini:EPL1999}.
We have calculated $\Deff(f)$ for ABPs within the stochastic simulations.
For small $\tauR$, the obtained behavior of $\Deff(f)$ is strikingly similar to that of $\mu(f)$ [\cref{fig:combo}(e,f)]; in particular, $\Deff(f)/\Dfree$ shows a peak near the transition ($f\approx \fc$), which grows in height without bounds as $\tauR \to 0$;
here, $\Dfree=\vA^2\tauR/3$ is the effective diffusion of the free ABP.
Such giant diffusion was studied for passive particles \cite{Reimann:PRL2001,*Reimann:PRE2002,LopezAlamilla:PRE2020,Lindner-Sokolov:PRE2016}
and has been seen in experiments \cite{Evstigneev-etal:PRE2008, Stoop-etal:NL2018};
a similar effect was unveiled recently for circle swimmers subject to gravity \cite{Chepizhko:PRL2022}.

In the lazy-wobbling limit (large $\tauR$), the corrugated potential induces also an enhanced dispersion.
In this regime, the corresponding data for $\Deff(f)/\Dfree$ depend only weakly on $\tauR$.
Invoking again the random walk picture of uncorrelated velocities $\vD(f;\theta_i)$ changing at a ``collision rate'' $\tauR^{-1}$ yields for the velocity autocorrelation function $Z(t) = \mathrm{Var}[\vD(f;\theta)]_{\unitvec u} \exp(-t/\tauR)$ \cite{Ehrenfest:1990}.
The Green--Kubo relation gives us
$\Deff(f) = \int_0^\infty Z(t) \diff t = \mathrm{Var}[\vD(f;\theta)]_{\unitvec u}  \tauR$;
the remaining $\unitvec u$-average is an elementary integral.
The lengthy result is given in Eqs.~(S21) and (S24) of \cite{supplement} and drawn in \cref{fig:combo}(f) (orange line), which shows that $\Deff(f)/\Dfree$ is maximal near $f \approx (\fc+\fc^+)/2$.
Expanding $D_\mathrm{max} \approx \Deff\boldsymbol((\fc+\fc^+)/2\boldsymbol)$ for $\vA \ll \vL$ yields
\begin{equation}
   D_\mathrm{max} \simeq (9\Dfree/8) (\vL/\vA + 3/5) ,
\end{equation}
which predicts an $6.3$-fold enhancement of $\Deff(f)$ over $\Dfree$ for $\vA = 0.2\vL$, as is observed in the data for $\tauR=10\tauL$ near $f\approx 1.1\fL$ [\cref{fig:combo}(f)].
We anticipate an arbitrarily large enhancement of the dispersion, $D_\mathrm{max} / \Dfree \sim 1/\vA$,  for weakly self-propelled particles with $\tauR \gtrsim \tauL$.

\paragraph{Conclusions.}

We have shown that activity impacts the depinning transition as follows:
the threshold force is shifted from its value $\fc$ for passive particles to $\fc^- < \fc$,
which depends on the propulsion strength $\vA$ but not on the persistence time $\tauR$ of rotational motion.
A sharp transition is preserved in the presence of active noise, in contrast to the rounding due to translational thermal noise. However, the approach to the transition point from above depends on $\tauR$ and the dimension $d$ of rotational Brownian motion: it obeys different power laws for the limits of the hyper and lazy wobbler with exponents $\beta=1/2$ (small $\tauR$) and $\beta' = d/2$ (large $\tauR$), respectively.
In between, $\vD(f)$ vanishes superexponentially fast, contrasting from the scenario of a $\tauR$-dependent exponent.
For the lazy wobbler, another singular point $\fc^+$ emerges as the mirror image of $\fc^-$ relative to $\fc$,
where the differential mobility $\mu(f)$ is maximum.
Concomitantly, the dispersion coefficient shows a giant enhancement, whose position depends on $\tauR$.
Overall, this qualitative change of the phenomenology is likely beyond the scope of a perturbative treatment of the passive case with $\tauR$ as the small parameter (e.g., \cite{Fodor:PRL2016,Szamel:PRE2014}).
Our work suggests further that probing nonlinear responses \cite{Winter:JCP2013,Senbil:PRL2019,Gruber:PRE2020} can contribute to a similar debate for arrested active matter \cite{Paul:PNAS2023,Debets:JCP2022,*Janssen:JPCM2019,Klongvessa:PRL2019}.

Our predictions appear amenable to experimental tests, e.g., using active colloidal particles
driven by external fields (e.g., gravitational \cite{Palacci:PRL2010, Ma-etal:SM2015, Thorneywork:PRL2017} or magnetic \cite{Tierno:PRL2012, Stoop-etal:PRL2020}) over a periodic landscape \cite{Wen:PRE2023, Choudhury:NJP2017, Gibbs:L2019},
and potentially for the chemotaxis of bacteria crawling on structured substrates \cite{Amselem:PO2012,Russbach:BJ2022}.
Experiments on active colloidal monolayers may give insight into the activity-induced depinning of collective variables, and our study is relevant for the melting transition of active colloidal crystals \cite{MassanaCid:2024}.
We also note that the lazy wobbler resembles a run-and-tumble motion with switching rate $\tau_R^{-1}$, which describes the motion of, e.g., \emph{E. coli} bacteria \cite{Schnitzer:PRE1993,Solon:EPJST2015}.

Finally, the active noise $\vA \unitvec u(t)$ differs qualitatively from the thermal, white noise $\vec \xi(t)$, both entering \cref{eqn:LE-gen-r}:
$\vA \unitvec u(t)$ is bounded in magnitude, but $\vec \xi(t)$ can assume arbitrarily large values. Only in the latter case, the probability to surmount the potential barrier is nonzero for any, even small driving force $f\geq 0$.
Second, the integral $\int_0^t \vA \unitvec u(s) \,\diff s$ is a finite-variation process, unlike the Wiener process $\int_0^t \vec \xi(s) \,\diff s$, and hence yields a drift rather than a diffusion term in the corresponding Fokker--Planck operator (also see \cite{supplement,Szamel:PRE2014}).
The active noise may thus be interpreted as a random tilting of the potential landscape but not as an intrinsic diffusion.
We anticipate that our findings go well beyond the active matter context and apply to any system with a saddle-node bifurcation in the presence of a bounded noise.

\begin{acknowledgments}
We thank Arkady Pikovsky for helpful discussions.
Financial support by Deutsche Forschungsgemeinschaft (DFG, German Research Foundation) under Germany’s Excellence Strategy---MATH+: The Berlin Mathematics Research Center (EXC-2046/1)---Project No.\ 390685689 (Subprojects AA1-18 and EF4-10)
and further under Project No.\ 523950429 is gratefully acknowledged.
\end{acknowledgments}

\bibliography{manuscript}

\clearpage
\onecolumngrid
\makeatletter\floats@sw{
\renewcommand{\theequation}{S\arabic{equation}}
\renewcommand{\thefigure}{S\arabic{figure}}
\renewcommand{\thetable}{S\arabic{table}}
\setcounter{equation}{0}
\setcounter{figure}{0}
\setcounter{table}{0}
}{}
\makeatother

\section*{Supplemental Material\\[-1mm]}

\section{Numerics of the active Brownian particle (ABP) model\\[-1mm]}

\subsection{Stochastic simulation of the It{\=o}--Langevin equations}

For the stochastic simulation of the ABP model given by Eqs.~(2) and (3) in the main text, we have generated, for each force $f$, up to \num{5e5} random trajectories $x(t)$ of length $\num{5e4}\tauL$.
To this end, we combined the Euler(--Maruyama) integration for the translational motion and a geometric integration scheme \cite{Hoefling:2024} for the rotational Brownian motion, using an integration time step of $\Delta t = \num{e-3} \tauL$.
In addition, we have applied a simple antithetic variance reduction technique, where for every noise realization $\vec\omega(t)$ one obtains two trajectories: one with $\vec\omega(t)$ and one with $-\vec\omega(t)$, exploiting the inflection symmetry of the noise.

\subsection{Numerical solution of the Fokker--Planck equation}

The Fokker--Planck equation (FPE) corresponding to the Itō--Langevin Eqs. (1) and (2) of the main text reads
\begin{equation}
  \partial_t p(\vec r,\vec u,t) =
    - \nabla \cdot [\mu_0 (\vec f - \nabla U(\vec r)) + \vA \unitvec u] \, p(\vec r,\vec u,t)
    + \tauR^{-1} L_{\unitvec u} \, p(\vec r,\vec u,t) \,,
  \label{eq:FP-general}
\end{equation}
where $p(x,\vec u,t)$ is the joint probability density of the position $x$ and the orientation $\unitvec u$ at time $t$ and $L_{\unitvec u}$ denotes the Laplace--Beltrami operator on the $d$-dimensional unit sphere.
For the one-dimensional corrugated potential landscape discussed in this work, only the projection $x = \vec r \cdot \unitvec e_x$ and the polar angle $\theta$ such that $z:=\cos(\theta) = \unitvec u \cdot \unitvec e_x$ are relevant.
Then, \cref{eq:FP-general} reduces to
\begin{equation}
  \partial_t p(x,z,t) =
    -\partial_x [\mu_0 f - \vL \sin(kx) + \vA z] \, p(x,z,t)
    + \tauR^{-1} \partial_z \mleft(1-z^2\mright) \partial_z \, p(x,z,t) \,,
  \label{eq:FP-1d}
\end{equation}
which is the FPE corresponding to Eqs. (2) and (3) of the main text.
The domain of $p(x,z,t)$ is $x \in \mathbb{R}$, $z \in [-1,1]$, and $t \geq 0$.

Exploiting the inherent $x$-periodicity of the problem, we proceed to the reduced probability density \cite{Reimann:PR2002}
$\hat p(x,z,t):=\sum_{n=-\infty}^{\infty} p(x+n\lambda,z,t)$, which satisfies \cref{eq:FP-1d} for $x \in [0,\lambda]$ with periodic boundary conditions, $\hat p(x,z,t) = \hat p(x+\lambda,z,t)$ with $\lambda=2\pi/k$.
We recall further that the eigenfunctions of the $d=3$ rotational diffusion operator are the Legendre polynomials $P_\ell(z)$, i.e.,
\begin{equation}
  \partial_z \mleft(1-z^2\mright) \partial_z P_\ell(z) = - \ell(\ell + 1) P_\ell(z) \,; \qquad \ell \in \mathbb{N}_0 \,.
  \label{eq:eigfun-P}
\end{equation}
The periodicity of $\hat p(x,z,t)$ in $x$ together with \cref{eq:eigfun-P} suggest to represent the solution as a Fourier--Legendre series,
\begin{equation}
  \hat p(x,z,t) = \sum_{n\in \mathbb{Z}} \sum_{\ell \geq 0} c_{n\ell}(t) \e^{\i n k x} P_\ell(z) \,.
  \label{eq:expansion}
\end{equation}
The time evolution of the coefficients $c_{n\ell}(t)$ is implied by \cref{eq:FP-1d} and one finds:
\begin{subequations}
\label{eq:evo-coeff}
\begin{align}
\dot c_{n0}  = &
  -\i n k \left[ \mu_0 f c_{n0} - \frac{\vL}{2\i} (c_{n-1,0} - c_{n+1,0}) \right]
  -\i n k \vA \frac{c_{n1}}{3} \,; \qquad \ell =0, \label{eq:evo-coeff-1} \\
\dot c_{n\ell}  = &
  -\i n k \left[ \mu_0 f c_{n\ell} - \frac{\vL}{2\i} (c_{n-1,\ell} - c_{n+1,\ell}) \right] \nonumber \\
  &-\i n k \vA \left(\frac{\ell}{2\ell-1} c_{n,\ell-1} + \frac{\ell+1}{2\ell+3} c_{n,\ell+1}  \right)
  -\tauR^{-1} \ell(\ell +1) c_{n\ell} \,; \qquad \ell > 0. \label{eq:evo-coeff-2}
\end{align}
\end{subequations}
For the stationary solution, the left hand sides are set to zero, $\dot c_{n\ell} = 0$, and
\cref{eq:evo-coeff} becomes a linear system in the coefficients $c_{n\ell}$.
The normalization condition $\int \hat p(x,z,t) \, \diff x \diff z = 1$ implies $c_{00}=1/2$, which renders the linear system inhomogeneous. We truncated the series \eqref{eq:expansion} symmetrically to keep only terms with $-N \leq n \leq N$ and $0 \leq \ell \leq L$ and solved the system of $(2N+1)\times(L+1)$ equations numerically using standard BLAS routines.

\begin{figure}
 \includegraphics[width=0.5\textwidth]{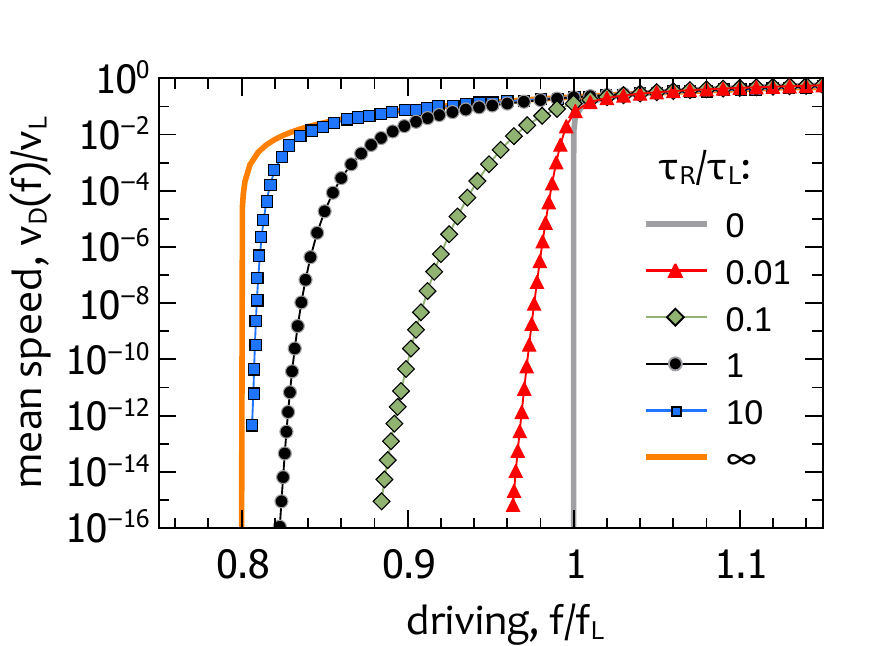}
 \caption{Drift velocity $\vD(f)$ as function of the driving force $f$ was obtained from the numerical FPE solution [\cref{eq:supp-mean-speed}] with the self-propulsion velocity fixed to $\vA=0.2\vL$. The same data are shown in Fig.~1c of the main text on a super-logarithmic scale.}
 \label{fig:PFE-numerics}
\end{figure}

The mean speed $\vD(f)=\lim_{t\to\infty}\langle \dot x(t) \rangle$ is the integral of the $x$-component of the probability flux
$\langle \dot x(t) \rangle = \int \hat\jmath_x(x,z,t) \,\diff x \diff z$
with $\hat\jmath_x(x,z,t) = (\mu_0 f-\vL \sin(kx) + \vA z) \hat p(x,z,t)$
and, upon using \cref{eq:expansion}, it is calculated from the expansion coefficients as
\begin{equation}
  \vD(f) = \mu_0 f + 2 \vL \Imag c_{1,0} + \frac{2}{3}\vA c_{0,1} \,.
  \label{eq:supp-mean-speed}
\end{equation}
The numerical results shown in \cref{fig:PFE-numerics} and in Fig.~1(c) of the main text
were obtained for $N=\num{10000}$ and $L=30$.
The different orders of magnitude for $N$ and $L$ were chosen to account for the observation that the eigenvalues of $\nabla$ scale as $n$, whereas the eigenvalues of $L_{\unitvec u}$ scale as $\ell(\ell+1)$.

\section{Drift velocity and dispersion coefficient of lazy wobblers}

\subsection{Random walk model}

As described in the main text, the trajectories $x(t)$ in the regime of the lazy wobbler ($\tauR \gg \tauL$ and $\tau_R \gg \tau_f$)  are approximated by a one-dimensional random walk (or ``flight'') such that the orientation of the particle changes instantaneously at random times with a rate $\tauR^{-1}$.
In this heuristic model, the orientation $\vec u(t)$ consists of piecewise constant segments $\vec u_i$ of random durations $\tau_i$ for $i=1,2,\dots$.
For the depinning problem, we may equivalently use the angles $\theta_i$ such that $\cos \theta_i = \unitvec u_i \cdot \unitvec e_x$.
Given a fixed orientation $\unitvec u_i$ (or $\theta_i$), the particle moves at the velocity $\vD(f;\theta_i)$ for a time span $\tau_i$.
Then, assuming $x(0)=0$, the spatial displacement after time $t$ is
\begin{equation}
  x(t) = \sum_{i=1}^{N(t)} \vD(f;\theta_i) \tau_i \,,
  \label{eq:supp-random-flight}
\end{equation}
where $N(t)$ is counting the reorientation events up to and including time~$t = \sum_{i=1}^{N(t)} \tau_i$.
The resulting trajectories $x(t)$ correspond exactly to the motion of run-and-tumble particles.

Following the ideas of Boltzmann's Stoßzahlansatz (molecular chaos hypothesis) \cite{Ehrenfest:1990}, the reorientation events (``collisions'') are assumed to be independent and combine exponentially distributed times $\tau_i$ between subsequent collisions with orientations $\vec u_i$ that are sampled independently from the equilibrium distribution, i.e., a uniform distribution on the unit sphere, $|\vec u|=1$.
As a consequence, $N(t)$ is a Poisson process with parameter $\tauR^{-1}$.

\subsection{Drift velocity}

For the drift velocity (or: mean speed), one finds from \cref{eq:supp-random-flight}: 
\begin{align}
 \vDa(f) &= \lim_{t\to \infty} \frac{x(t)}{t} \notag \\
  &= \lim_{N \to \infty} \frac{1}{N}\sum_{i=1}^N \vD(f;\theta_i) \tau_i  \bigg/ \frac{1}{N}\sum_{i=1}^N \tau_i \notag \\
  &= \frac{\langle \vD(f;\theta_i) \rangle_{\vec u} \langle \tau_i \rangle}{\langle \tau_i \rangle} \notag \\
  &= \frac{1}{4\pi} \int \vD(f;\theta) \sin(\theta) \,\diff \theta \diff \phi \,.
  \label{eq:vDa-S1}
\end{align}
In the second line, we have used that $N(t\to \infty) \to \infty$ monotonically, which permits that the limit $t\to \infty$ is replaced by letting $N \to \infty$.
The third line follows from the strong law of large numbers and the independence of $\theta_i$ and $\tau_i$.
The last line of \cref{eq:vDa-S1} represents the orientation-averaged drift velocity, $\langle \vD(f;\theta_i) \rangle_{\vec u}$.
We rewrite the integrand, as in the main text, in terms of $\vD(f;\theta) = \vL s(\fA(\theta)/\fL)$ with the effective driving force $\fA(\theta)=f+(\vA/\mu_0)\cos\theta$ and $s(y) = \sqrt{y^2 - 1}$ for $|y| > 1$ and $s(y)=0$ otherwise. Substituting $u_x = \cos \theta$, the $\vec u$-average is calculated as:
\begin{align}
  \vDa(f) &= \frac{\vL}{2} \int_{-1}^1 s\bigl(f/\fL+(\vA/\vL) u_x\bigr) \,\diff u_x \notag \\
  &= \frac{\vL^2}{2\vA} \int\limits_{\hspace{2.3em}\mathclap{\max(y_-, 1)}}^{\hspace{2.3em}\mathclap{\max(y_+, 1)}} s(y) \,\diff y \,,
  \label{eq:S-drift-integral}
\end{align}
after substituting $y = \fA(\theta)/\fL = f/\fL + (\vA/\vL) u_x$ for $u_x$.
The integral bounds $y_{\pm}=f/\fL \pm \vA/\vL$ have been tightened to the condition $|y| > 1$, where the integrand is nonzero.
The remaining integral is elementary:
\begin{equation}
  w(y) := \int_1^y s(y') \diff y' = \frac{1}{2}[y s(y) - \ln\boldsymbol(y + s(y)\boldsymbol)] \,.
  \label{eq:w-fct}
\end{equation}
Introducing $w_{\pm}(f/\fL):= w(y_\pm) = w(y \pm \vA/\vL)$ and noting that $w_-(\fc^+/\fL) = w(1) = 0$, we obtain the result quoted in Eq.~(6) of the main text:
\begin{align}
  \vDa(f) = \frac{\vL^2}{2 \vA}
    \begin{cases}
    0, & f \leq \fc^-, \\
    w_{+}(f/\fL), & \fc^- < f <  \fc^+, \\
    w_{+}(f/\fL) - w_{-}(f/\fL), & f \geq \fc^+ .
    \end{cases}
  \label{eq:supp-vDa}
\end{align}

\subsection{Critical behavior}

To obtain the critical behavior of the drift velocity, we introduce the distance to the critical point, $\epsilon = (f-\fc^-)/\fL$, and find the leading term in an asymptotic expansion of the integral in \cref{eq:S-drift-integral}.
Restricting to $0 < \epsilon < 2\vA/\vL$, it holds $y_+=1+\epsilon$ and $y_-=1+\epsilon-2\vA/\vL < 1$, which simplifies the integral bounds.
Introducing a new integration variable $0 \leq \eta \leq 1$ such that $y = 1 + \eta \epsilon$ yields:
\begin{align}
\vDa(f = \fc^- + \epsilon \fL)
  &= \frac{\vL^2}{2\vA} \int\limits_1^{1 + \epsilon} s(y) \,\diff y
  = \frac{\vL^2}{2\vA} \, \epsilon \int\limits_0^1 \sqrt{2\eta \epsilon} \, [1 + O(\epsilon)] \,\diff \eta  \notag \\
  &= \frac{\vL^2}{2\vA} \frac{2\sqrt{2}}{3} \, \epsilon^{3/2}  \, [1 + O(\epsilon)] \,.
\end{align}
We used that $s(y)$ is bounded on the domain of integration, which permits interchanging the $\eta$-integral with the
expansion for $\epsilon \to 0$.
Hence,
\begin{equation}
  \vDa(f \downarrow \fc^-)
    \simeq \frac{\sqrt{2}}{3} \frac{\vL^2}{\vA} \, \epsilon^{3/2}
    \sim (f - \fc^-)^{3/2} \,.
  \label{eq:S-vDa-crit3D}
\end{equation}
Alternatively, the same result is obtained by expanding $w_+(1+\epsilon)$ defined after \cref{eq:w-fct}.

\subsection{Extension to rotational motion in the plane}

The preceding analysis of the lazy-wobbling limit has a straightforward extension to ABP models with two-dimensional rotational motion, where the self-propulsion velocity is constrained to the plane of translational motion.
The essential difference is that the orientation vector $\vec u$ is uniformly distributed on a circle rather than on a sphere,
which has implications for the integrals implementing the $\vec u$-average.
For the mean speed, \cref{eq:vDa-S1} is replaced by
\begin{equation}
 \vDa(f) = \frac{1}{\pi} \int_{0}^\pi \vD(f;\theta) \,\diff \theta \,, \label{eq:vDa-2Dint}
\end{equation}
where we stick to a representation in terms of the polar angle $\theta \in [0,\pi]$.
Relative to \cref{eq:vDa-S1}, the factor $\sin(\theta)$ is missing from
the differential of the solid angle.
Nevertheless, we substitute $u_x = \cos(\theta)$ with $\diff u_x = \sin(\theta) \,\diff \theta = \sqrt{1 - u_x^2} \, \diff \theta$ and, subsequently, introduce $y$ as above. With this, the expression corresponding to \cref{eq:S-drift-integral} reads
\begin{align}
  \vDa(f) &= \frac{\vL}{\pi} \int_{-1}^1 \frac{s\bigl(f/\fL+(\vA/\vL) u_x\bigr)}{\sqrt{1 - u_x^2}} \,\diff u_x \notag \\
    &= \frac{\vL^2}{\pi\vA} \;\:\int\limits_{\hspace{-2em}\vphantom{\bigl(}\mathrlap{\max(y_-, 1)}}^{\hspace{-2em}\vphantom{\bigl(}\mathrlap{\max(y_+, 1)}} \hspace{1.5em}
    \frac{s(y)}{\sqrt{1 - r(y)^2}} \,\diff y \,,
  \label{eq:S-drift-integral-2D}
\end{align}
upon replacing $u_x = r(y) := (\vL/\vA)(y - f/\fL)$ by $y$ and $y_{\pm}=f/\fL \pm \vA/\vL$, as before.

In the absence of an explicit form for the integral in \cref{eq:S-drift-integral-2D}, we determine the critical behavior close to the critical point analogously as above for $d=3$.
Writing again $f = \fc^- + \epsilon \fL$,
it holds $r(y; \epsilon)=1 + (\vL/\vA)(y-1-\epsilon)$.
For $0 < \epsilon < 2\vA/\vL$, we thus have
\begin{equation}
  \vDa(f = \fc^- + \epsilon \fL)
    = \frac{\vL^2}{\pi\vA} \:\int\limits_{\hspace{0 em}\mathrlap{1}}^{\hspace{-1.5em}\mathrlap{1+\epsilon}} \quad
    \frac{s(y)}{\sqrt{1 - r(y; \epsilon)^2}} \,\diff y \,.
  \label{eq:S-drift-integral-2D-2}
\end{equation}
Passing on to the integration variable $\eta$ such that $y = 1+\eta \epsilon$, the leading order in $\epsilon$ is obtained by letting $\epsilon \to 0$ in the integrand:
\begin{align}
  \vDa(f = \fc^- + \epsilon \fL)
  &= \frac{\vL^2}{\pi\vA} \int_0^1
    \frac{\sqrt{2\eta \epsilon  + O(\epsilon^2)}}{\sqrt{2(\vL/\vA) (1-\eta) \epsilon + O(\epsilon^2)}}
    \, \epsilon \diff \eta  \notag \\
  &\simeq \frac{\vL^{3/2} \epsilon}{\pi\vA^{1/2}} \int_0^1 \sqrt{\eta / (1-\eta)} \, \diff \eta \notag \\
  &= \frac{\vL^{3/2} \epsilon}{2 \vA^{1/2}} \,;
\end{align}
the integral in the last step evaluates to $\pi/2$.
Thus close to the critical point, it holds for the $d=2$ case:
\begin{equation}
  \vDa(f \downarrow \fc^-) \simeq \frac{1}{2} (\vL/\vA)^{1/2} \mu_0 (f-\fc^-)
    \sim f-\fc^- \,.
  \label{eq:S-vDa-crit2D}
\end{equation}

\begin{figure}
 \includegraphics[width=0.55\textwidth]{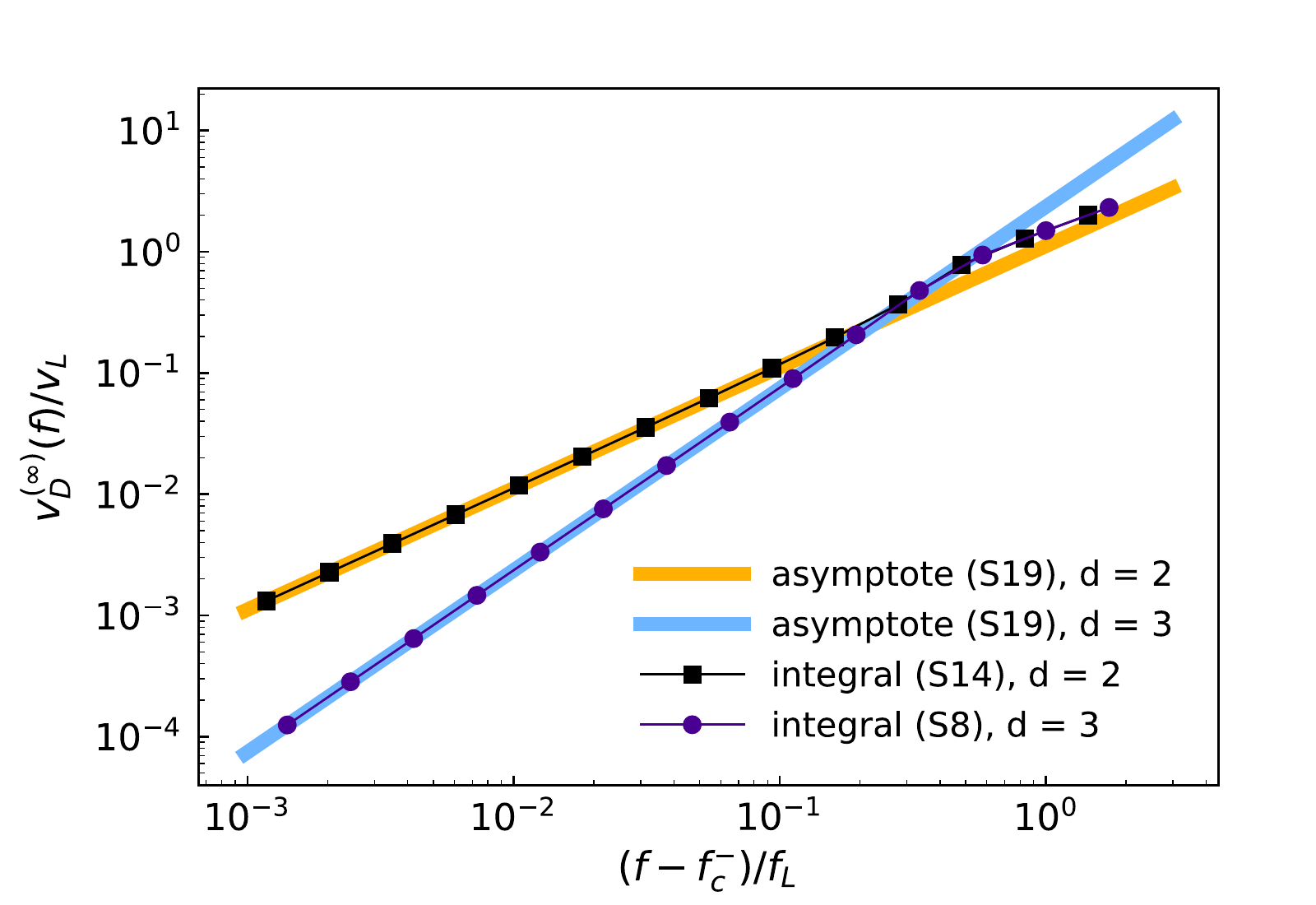}
 \caption{Critical behavior of the drift velocity $\vDa(f)$ in the lazy-wobbling limit ($\tauR\to\infty$) as function of the distance to the critical point, $\epsilon = (f-\fc^-)/\fL$, evaluated for $\vA/\vL=0.2$ and for rotational motion in $d=2$ and $d=3$ dimensions.
 Solid lines show the analytic prediction for both cases [\cref{eq:S-vDa-critdD}]
 and symbols denote numerical results from the quadrature of the orientational $\vec u$-average
 given in \cref{eq:vDa-2Dint} for $d=2$ (squares) and \cref{eq:vDa-S1} for $d=3$ (circles);
 the latter agree also with the explicit expression in \cref{eq:supp-vDa}.
 }
 \label{fig:crit-behav}
\end{figure}

The critical laws in \cref{eq:S-vDa-crit3D,eq:S-vDa-crit2D} can be summarized for $d=2, 3$ as
\begin{equation}
  \vDa(f \downarrow \fc^-) \simeq \frac{1}{d} \bigl[(d-1) \vL \vA\bigr]^{1/2} \, \bigl[\mu_0 (f-\fc^-)/\vA\bigr]^{d/2} \,.
  \label{eq:S-vDa-critdD}
\end{equation}
\Cref{fig:crit-behav} corroborates this analytic result, which coincides asymptotically ($f \downarrow \fc^-$) with the data for $\vDa(f)$ from the quadrature of the $\vec u$-average [\cref{eq:vDa-S1} for $d=3$, \cref{eq:vDa-2Dint} for $d=2$].

\subsection{Dispersion coefficient}

Concerning the dispersion of the trajectories, we note first that the sequence of reorientations yields, in full analogy to the particle collisions in a dilute gas, for the velocity autocorrelation function:
\begin{align}
  Z(t) &= \langle [\vD(f;\theta(t)) - \vDa(f)] \, [\vD(f;\theta(0)) -\vDa(f)] \rangle \notag \\
  &= \mathrm{Var}[\vD(f;\theta_i)]_{\unitvec u} \, \e^{-t/\tauR} \,,
\end{align}
taking into account that $\langle \vD(f;\theta_i) \rangle_{\vec u} = \vDa(f)$ may not be zero.
The factor $\e^{-t/\tauR}$ is simply the probability that no ``collision'' has occurred in the time interval $[0,t]$.
The effective diffusion coefficient then follows from the Green--Kubo relation:
\begin{align}
 \Deff(f) &= \int_0^\infty Z(t) \, \diff t = \mathrm{Var}[\vD(f;\theta_i)]_{\unitvec u} \tauR \notag \\
 &= \bigl(\langle \vD(f;\theta_i)^2 \rangle_{\unitvec u} - \vDa(f)^2 \bigr) \tauR \,.
 \label{eq:supp-Deff}
\end{align}
It remains to compute the second moment, $\langle \vD(f;\theta_i)^2 \rangle_{\unitvec u}$.
The same arguments apply that have led to \cref{eq:S-drift-integral} for the first moment in the case $d=3$. Therefore:
\begin{align}
  \langle \vD(f;\theta_i)^2 \rangle_{\vec u}
  &= \frac{\vL^2}{2} \int_{-1}^1 s(f/\fL+(\vA/\vL) u_x)^2 \diff u_x \notag \\
  &= \frac{\vL^3}{2\vA} \: \int\limits_{\hspace{-1.5em}\mathrlap{\max(y_-, 1)}}^{\hspace{-1.5em}\mathrlap{\max(y_+, 1)}}
    \hspace{1.5em} \bigl(y^2 - 1\bigr) \,\diff y
\end{align}
Introducing
$
  \tilde w(y) := \int_1^y s(y')^2 \, \diff y' = \frac{1}{3}\bigl(y^3 - 1\bigr) + 1 - y
$
and $\tilde w_\pm(f/\fL) = \tilde w(y_\pm)$ and noting that $\tilde w(1) = 0$, it follows
\begin{align}
  \langle \vD(f;\theta_i)^2 \rangle_{\vec u} = \frac{\vL^3}{2 \vA}
    \begin{cases}
    0, & f \leq \fc^-, \\
    \tilde w_{+}(f/\fL), & \fc^- < f <  \fc^+, \\
    \tilde w_{+}(f/\fL) - \tilde w_{-}(f/\fL), & f \geq \fc^+ ,
    \end{cases}
\end{align}
which can be rewritten in the form
\begin{align}
  \langle \vD(f;\theta_i)^2 \rangle_{\vec u} &= \vL^2 \times \begin{cases}
    0,  & f \leq \fc^- , \\
    \displaystyle \frac{\vL/\vA}{6\fL^3} \, (f - \fc^-)^2 (f+\fc^+ + \fL), & \fc^- < f < \fc^+ , \\
    \displaystyle (f/\fL)^2 + \frac{1}{3} (\vA/\vL)^2 - 1, & f \geq \fc^+ .
  \end{cases}
  \label{eq:supp-vD2}
\end{align}
The dispersion coefficient $\Deff(f)$ is obtained by inserting \cref{eq:supp-vDa,eq:supp-vD2} into \cref{eq:supp-Deff}, and its behavior is exemplarily shown in Fig.~1(f) of the main text (orange line).

\addtocounter{NAT@ctr}{8}


\end{document}